\documentclass[12pt]{article}
\newcommand{\sect}[1]{\section{#1}\setcounter{equation}{0}}

\begin{document}
\begin{titlepage}

\begin{flushright}
UCSBTH-98-7 \\ 
NSF-ITP-98-093 \\
hep-th/9810200
\end{flushright}
\bigskip
  
\begin{center}
{\Large \bf Microcanonical Phases of String Theory on
AdS$_m\times$S$^n$}
\bigskip\bigskip
  
{\large Amanda W. Peet\footnote{\tt
peet@itp.ucsb.edu}}\bigskip

Institute for Theoretical Physics, \\
University of California,\\
Santa Barbara, CA 93106 \\
\bigskip\bigskip

{\large Simon F. Ross\footnote{\tt
 sross@cosmic.physics.ucsb.edu}} \bigskip

Department of Physics\\
University of California,\\
 Santa Barbara, CA 93106\\
  
\end{center}
\vskip 1.5cm
  
\begin{abstract}
Banks, Douglas, Horowitz and Martinec \cite{bdhm:adscft} recently
argued that in the microcanonical ensemble for string theory on
AdS$_m\times$S$^n$, there is a phase transition between a
black hole solution extended over the S$^n$ and a solution localized
on the S$^n$. If we think of this AdS$_m\times$S$^n$ geometry as
arising from the near-horizon limit of a black $m-2$ brane, the
existence of this phase transition is puzzling. We present a
resolution of this puzzle, and discuss its significance from the point
of view of the dual $m-1$ dimensional field theory. We also
discuss multi-black hole solutions in AdS.
\end{abstract} 
\end{titlepage}

\sect{Introduction}

The recently discovered AdS/CFT duality
\cite{juan:N1,witten:eucl,gub:corr} between string theory in the bulk
of anti-de Sitter spaces (times spheres) and large-$N$ conformal field
theories gives new insights into both the gauge theory and the nature
of the bulk theory. In an early application, Witten
\cite{witten:therm} used this duality to relate the thermodynamics of
asymptotically anti-de Sitter spaces \cite{hawk:adstherm} to the
expected thermodynamics of the gauge theory. Recently, Banks, Douglas,
Horowitz and Martinec \cite{bdhm:adscft} studied the microcanonical
ensemble to determine the spectrum of string theory on these
backgrounds in more detail. At high energies, the typical state is a
Schwarzschild-AdS black hole. They argued that at lower energies,
where the horizon radius is smaller than the cosmological scale, $r_+
< b$, the black hole will localize on the sphere due to the
Gregory-Laflamme instability\cite{greg:unstable}.  The typical state
at lower energies will then be a $D$ dimensional Schwarzschild black
hole.

Certain D$p$-branes have AdS$_{p+2}\times$S$^{D-p-2}$ spacetimes as
their near-horizon geometries: the D3-brane, D1+D5 system, M2, and
M5-branes have $(p,D)=(5,10), (1,6), (2,11), (5,11)$ respectively.
The existence of the above localization instability in the
near-horizon region should then imply some instability of these
D$p$-branes. However, as the S$^{D-p-2}$ corresponds to a sphere
surrounding the D$p$-brane, this is {\em not} the usual localization
instability in the direction along the brane. Rather, it would imply
that the stable solution is one in which the geometry is not
spherically symmetric. This conclusion runs counter to the black hole
no-hair theorems.  We would also expect that any such asphericity
would be radiated away. Thus there is apparently a puzzling
contradiction between the expectations from the near-horizon region
and the full asymptotically flat solution.

To understand the resolution of this puzzle, we consider the
thermodynamics in more detail, especially the question of how it is
affected by the asymptotic boundary conditions. In
\cite{witten:therm}, Witten considered two sets of asymptotic boundary
conditions. The conformal boundary was taken to be either $S^p \times
S^1$ or ${\rm\bf R}^p \times S^1$.\footnote{For pure AdS$_{p+2}$, the
boundary is $S^{p+1}$, which is conformally equivalent to
${\rm\bf R}^{p+1}$. $S^p \times S^1$ and ${\rm\bf R}^p \times S^1$ are,
however,
not conformally equivalent; for instance, the former has a conformally
invariant parameter, the ratio of the two radii, while the latter
does not.} The discussion in
\cite{bdhm:adscft} corresponds to the microcanonical version of the
former choice, whereas the near-horizon limit of a D$p$-brane
corresponds to the latter \cite{witten:therm,hor:resol}. If we
compactify the directions along the D$p$-brane, the conformal boundary
is $T^p \times S^1$.

In section \ref{sec2}, we consider the conformal branes, in particular
the D3-brane, for which the near-horizon limit is AdS$_5\times$S$^5$.
We show that for a conformal boundary where the spatial part is $T^3$
(or ${\bf R}^3$), there is no localization instability on the S$^{5}$
in the microcanonical ensemble. The discussion is entirely similar for
the M2- and M5-branes, and we state the results for these cases as
well.  We discuss the D1+D5-brane system, i.e., AdS$_3$, separately in
section \ref{sec3}, as in this case, the distinction between different
boundary conditions is more subtle. The spatial boundary is just a
circle, but the six dimensional Schwarzschild black hole could be
embedded in either an AdS$_3$ background, or in the $M=0$ BTZ black
hole.  We give a physical argument for preferring the latter.  In
section
\ref{sec4} we discuss multi-black hole solutions in AdS backgrounds,
and argue that the toroidal black holes cannot split up into
multi-black holes.  We conclude with a brief discussion. 

\sect{Localization on S$^{D-p-2}$ versus boundary conditions}
\label{sec2}

In this section, we consider the asymptotically
AdS$_{p+2}\times$S$^{D-p-2}$ spacetimes, which are related to D3-,
D1+D5-, M2- and M5-branes for $(p,D)=(3,10),(1,6),(2,11),(5,11)$
respectively.  We will begin by reviewing the case of a spacetime with
spherical boundary conditions, which was discussed in \cite{bdhm:adscft}
for $p=3,1$. For high energies, the dominant contribution comes from
the Schwarzschild-AdS black hole (times an S$^{D-p-2}$)
\cite{witten:therm}. The metric of the asymptotically AdS factor
is\footnote{We will write metrics in Euclidean signature, and use the
canonical ensemble as a ``trick'' to calculate the entropy, but our
physical interest is in the microcanonical ensemble.}
\begin{equation} \label{sads}
ds^2 = \left({{r^2}\over{b^2}} +1 - {{w_{p+2} M}\over{r^{p-1}}}\right)
d\tau^2 + \left({{r^2}\over{b^2}} +1 - {{w_{p+2}
M}\over{r^{p-1}}}\right)^{-1} dr^2 + r^2 d\Omega_p ,
\end{equation}
where $w_{p+2} = 16\pi G_{p+2}/[p {\rm Vol}(S^p)] \sim G_D/b^{D-p-2}$,
and $G_d$ is the $d$-dimensional Newton constant.  The radius $b$ of the 
AdS factor depends on the brane:
\begin{equation}\label{bdefs}
b_{D3} \sim (g_s N)^{1/4} \ell_s , \quad 
b_{M2} \sim \ell_{11} N^{1/6}, \quad 
b_{M5}  \sim \ell_{11} N^{1/3}, \quad 
b_{D1+D5} \sim (g_6 N_1 N_5)^{1/4} \ell_s,
\end{equation}
where $\ell_s$ is the string length and $\ell_{11}=\ell_s g_s^{1/3}$
is the eleven-dimensional Planck length. The event horizon of this
asymptotically AdS black hole is at $r=r_+$, where $r_+$ solves the
equation
\begin{equation} \label{srp}
{{r_+^2}\over{b^2}} +1 - {{w_{p+2} M}\over{r_+^{p-1}}} = 0. 
\end{equation}
As shown in \cite{hawk:adstherm}, the entropy is
given by an expression familiar from asymptotically flat spaces,
\begin{equation} \label{sent}
S = {1 \over 4G_{p+2}} r_+^p {\rm Vol}(S^p). 
\end{equation}
There is no elementary expression for the entropy as a function of
mass. If we rewrite the horizon position relation (\ref{srp}) as an
expression for the mass, 
\begin{equation}
M = {{p{\rm Vol}(S^p)}\over{16\pi G_{p+2}}} 
\left({{r_+^{p+1}}\over{b^2}}+r_+^{p-1}\right),
\end{equation}
we see that there are two limits of the parameter $r_+/b$ in which
there is a simple approximate expression for the entropy.  Black holes
whose horizon is large by comparison to the radius of curvature of the
AdS$_{p+2}$ have $r_+/b \gg 1$.  In this case, the first term in the
mass dominates, and hence $S \sim
(b^{D-2}/G_D)\,(G_DM/b^{D-3})^{p/(p+1)}$.  For small black holes,
$r_+/b \ll 1$, the second term dominates, and so
\begin{equation} \label{entmass}
S \sim {1 \over G_{p+2}} \left(G_{p+2} M\right)^{p/(p-1)} =
{{b^p}\over{G_{p+2}}}\left({{G_{p+2}M}\over{b^{p-1}}}\right)^{p/(p-1)}
\sim 
{{b^{D-2}}\over{G_D}}\left({{G_DM}\over{b^{D-3}}}\right)^{p/(p-1)}.
\end{equation} 
In the canonical ensemble, these small black holes have negative
specific heat, and are unstable.  However, we work in the
microcanonical ensemble, where the energy (rather than the temperature
of the heat bath) is fixed and this instability is absent. Instead, a
different kind of instability is present.

On scales much less than the radius of curvature of the AdS$_{p+2}$,
the spacetime away from the black hole horizon looks approximately
like flat $D$-dimensional spacetime. The entropy $S'$ of
$D$-dimensional Schwarzschild black holes has a different dependence
on mass than that for the small Schwarzschild-AdS$_{p+2}$ black holes:
\begin{equation} \label{fsent}
S' \sim {{1}\over{G_D}} \left(G_DM\right)^{(D-2)/(D-3)} =
{{b^{D-2}}\over{G_D}}\left({{G_DM}\over{b^{D-3}}}\right)^{(D-2)/(D-3)}.
\end{equation} 
By comparing this with (\ref{entmass}), we see that the small
Schwarzschild-AdS black holes are entropically unstable, and will
undergo a localization transition leading to a $D$-dimensional
Schwarzschild black hole in this approximately flat region.  The
crossover happens when $r_+/b \sim 1$, and the entropy of the
Schwarzschild black hole is larger for $r_+/b < 1$.

These properties of Schwarzschild-AdS black holes are of course all
dependent on the form of the metric, which is in turn crucially
dependent on the boundary conditions.  We now turn to near-horizon
limits of conformal branes in order to see if the puzzle persists in
these geometries. The dominant contribution to the physics at high
energies will come from a black hole with a toroidal horizon
\cite{mann:top,hor:resol,witten:therm}. This black hole metric is
obtained directly as the near-horizon limit of the brane metric.  For
the purposes of illustration we will specialize to the D3-brane case,
and comment on the M-branes at the end of the section.  The toroidal
black hole metric is then
\begin{equation} \label{tor}
ds^2 = {\ell_s^4 U^2 \over b^2} 
\left[ \left( 1 - {U_0^4 \over U^4} \right) d\tau^2 
+ dy^i dy_i \right] + {b^2 \over U^2} \left( 1 - {U_0^4 \over
U^4} \right)^{-1} dU^2.
\end{equation}
We take the $y_i$ to be periodic, $y_i \equiv y_i +L$. We can of
course change the periodicity in $y_i$ and the value of $U_0$ by a
coordinate transformation, but the combination
\begin{equation}\label{rho0d3}
\rho_{0\,(D3)}\equiv {{\ell_s^2 U_0 L}\over{b}}
\end{equation}
is coordinate-invariant. This is the proper length of the compactified
directions at the event horizon.  

To calculate the entropy, we use the canonical ensemble. The
conformal boundary is $T^3 \times S^1$, rather than $S^3 \times S^1$,
and $\rho_{0}$ determines the ratio of the size of the $T^3$ to the
size $\beta$ of the $S^1$, which is the only conformally invariant
boundary datum.  In particular, in direct analogy with the analysis
\cite{witten:therm} of the $S^3$ case, we calculate the entropy
by varying with respect to the invariant quantity
\begin{equation}\label{gamrho0}
\gamma \equiv {{b\,\beta}\over{L}} = {{\pi b^2}\over{\rho_0}}
\end{equation}
rather than the temperature $\beta$.  The evaluation of the action
follows the same lines as in \cite{hawk:adstherm}.  We find
\begin{equation} \label{action}
I = -\  {1 \over 16 G_5} \rho_0^3 .
\end{equation}
As in \cite{witten:therm}, we use this action to approximate the
partition function, and obtain the mass and then the entropy by
varying with respect to $\gamma$. The mass of this toroidal black hole
is then
\begin{equation} \label{t3m}
M = {{1}\over{b}} {{3 b^3}\over{16\pi G_5}} 
\left({{\rho_0}\over{b}}\right)^4 
\sim {{b^7}\over{G_{10}}} \left({{\rho_0}\over{b}}\right)^4 .
\end{equation}
Note that this differs from the excess energy over extremality by a
dimensionless factor $(L/b)$.  The entropy is then calculated via
\begin{equation}
S = \gamma M - I =
\left(\gamma{{\partial}\over{\partial\gamma}}-1\right) I,
\end{equation}
which yields
\begin{equation} \label{tent}
S = {{b^3}\over{4G_5}} \left({\rho_0\over{b}}\right)^3
\sim {{b^8}\over{G_{10}}}\left({{G_{10}M}\over{b^7}}\right)^{3/4}.
\end{equation}

{}From this we see the essential difference between the toroidal and
spherical conformal boundaries: here the entropy as a function of mass
is the same power law, $S\sim M^{3/4}$, for all horizon sizes.
This difference results directly from the different form of the metric
(\ref{tor}) as compared to (\ref{sads}).  With our definition of mass,
it is also consistent with the observation in \cite{witten:therm} that
this black hole (without the periodic identifications,
i.e., $L\rightarrow\infty$) can be obtained from Schwarzschild-AdS by
taking the large-mass limit of the latter.  It also implies that the
specific heat of these toroidal black holes is
positive.

To see if there is a localization instability in the microcanonical
ensemble of the type found for the spherical boundary conditions, we
compare the toroidal solution to the $D=10$ Schwarzschild black hole.
Substituting $D=10$ into the expression (\ref{fsent}) for the
Schwarzschild entropy, we find
$S'\sim(b^8/G_{10})(G_{10}M/b^7)^{8/7}$.  This is comparable to the
toroidal black hole entropy (\ref{tent}) when $(G_{10}M/b^7)\sim{1}$,
and using the mass formula (\ref{t3m}) we see that this happens when
$\rho_0\sim{b}$, i.e., when the horizon size is of order the
cosmological scale. But while the entropy of the $D=10$ Schwarzschild
black hole varies more slowly with energy (mass) than
Schwarzschild-AdS, it varies more rapidly than the entropy of the
toroidal black hole. Therefore, even though the entropies agree when
the horizon size is of order the cosmological scale, the $D=10$
Schwarzschild entropy is lower at smaller energies and so there is
{\em no} localization instability.  (This also applies for ${\bf R}^3$,
i.e., the $L\rightarrow\infty$ limit.)  Of course, the fact that the
$D=10$ Schwarzschild entropy is larger for larger black holes does not
make the large toroidal black holes entropically unstable either,
because the transition to a $D=10$ Schwarzschild black hole was
possible only for black holes smaller than $b$, i.e., where we could
not ``see'' the cosmological constant.

Although we have explicitly analyzed only the D3-brane, we can easily
extend this to the M2- and M5-branes.  In the case of the M2-brane,
the nonextremality is parameterized by the function $(1-U_0^3/U^3)$,
because the variable $U$ in which the asymptotic AdS$_4\times$S$^7$
structure is manifest is related to the radial variable $r$ by
$r=U^{1/2}\ell_{11}^{3/2}$, rather than the more familiar D-brane
relation $U=r/\ell_s^2$.  In this case, the proper size of the horizon
is given by
\begin{equation}
\rho_{0\,(M2)} \sim {{\ell_{11}^3 U_0 L}\over{b^2}} .
\end{equation}
For the M5-brane, nonextremality is parameterized by the function
$(1-U_0^6/U^6)$, because $r=U^2\ell_{11}^3$, and 
\begin{equation}
\rho_{0\,(M5)} \sim {{\ell_{11}^{3/2} U_0 L}\over{b^{1/2}}} .
\end{equation}
For both M-branes, the inverse temperature scales as
$\beta\sim\sqrt{N}/U_0$, and the conformally invariant boundary datum
scales as $\gamma=\beta(b/L)\sim b^2/\rho_0$, as was the case for the
D3-brane in (\ref{gamrho0}).  Then using the equations (\ref{bdefs})
for the radius $b$ of the AdS, we find that the mass scales as
$M\sim(1/b)(b^p/G_{p+2})(\rho_0/b)^{p+1}\sim(b^8/G_{11})(\rho_0/b)^{p+1}$,
and the entropy as
$S\sim(b^p/G_{p+2})(\rho_0/b)^p\sim(b^9/G_{11})(G_{11}M/b^8)^{p/(p+1)}$.
Again, by comparing with the Schwarzschild entropy (\ref{fsent}),
which for $D=11$ scales as $S'\sim (b^9/G_{11})(G_{11}M/b^8)^{9/8}$,
we see that there is {\em no} S$^{D-p-2}$ localization instability
with a toroidal boundary, because the entropy of these small toroidal
black holes dominates that of the eleven dimensional Schwarzschild
black holes.

\sect{AdS$_3$ and two notions of mass}
\label{sec3}

For the AdS$_3\times$S$^3$ spacetimes, which arise in the near-horizon
limit of the D1+D5-brane system, the spherical and toroidal boundary
conditions degenerate to a single case, where the spatial boundary is
just a circle.  The dominant contribution at high energies comes from
the BTZ black hole \cite{ban:2+1},
\begin{equation} \label{BTZ}
ds^2 = (r^2/b^2 - G_3 M) d\tau^2 + {dr^2 \over (r^2/b^2-G_3 M)} + r^2
d\phi^2,
\end{equation}
where $M$ is the ADM mass, and the entropy is
\begin{equation} \label{3S}
S = {\pi r_+ \over 2 G_3} = {\pi b \sqrt{G_3 M} \over 2 G_3} \sim
{b^{5/2} \sqrt{G_6 M} \over G_6}.
\end{equation}
Pure AdS$_3$ is given by the BTZ black hole (\ref{BTZ}) with $G_3
M=-1$. In the Euclidean approach to the calculation of this entropy,
the action is calculated using the $M=0$ black hole as a background. 

We want to compare this to the entropy for a $D=6$ Schwarzschild black
hole, which is
\begin{equation} \label{6S}
S' \sim {{1}\over{G_6}} (M' G_6)^{4/3} .
\end{equation}
If we consider this black hole inserted into a pure AdS$_3$ ($\times
$S$^3$)
background, then we should take $M' = M + 1/G_3$, so that the ADM mass
conjugate to time $t$ is $M$. Alternatively, if the background
geometry should be the $M=0$ BTZ black hole ($\times
$S$^3$), then $M' = M$. In
\cite{bdhm:adscft}, the former alternative was implicitly taken. There
is then a localization transition at $M \sim 1/G_3$ between the BTZ
black hole and a $D=6$ Schwarzschild black hole embedded in AdS$_3$.

The BTZ black hole (\ref{BTZ}) is the near-horizon limit of a
compactified black string; the compactified direction along the string
becomes the angular direction in the BTZ solution. Pure AdS$_3$ cannot
be obtained as the near-horizon limit of some regular string
solution\footnote{In this solution, the direction along the string must
also be compact, if the compactified black string is to decay into
it. AdS$_3$ is of course the near-horizon limit if this direction is
not compactified.}; the mass parameter $M$ in (\ref{BTZ}) is
proportional to the energy above extremality of the string, which
cannot be negative. We therefore argue that if we are considering this
$M>0$ BTZ spacetime as the near-horizon limit of a D1+D5-brane system,
then we should compare to a $D=6$ Schwarzschild black hole in an $M=0$
BTZ background, not the AdS$_3$ background. In this case, the
entropies are still comparable when $M \sim 1/G_3$, but as we lower
$M$, the entropy of the $D=6$ Schwarzschild black hole decreases more
quickly than that of the BTZ black hole. Therefore, in the
near-horizon limit of the D1+D5-brane system, there is {\em no}
localization instability on S$^3$.

\sect{Multi-black hole instability}
\label{sec4}

The black hole solutions (\ref{tor},\ref{BTZ}) which appear when the
boundary at infinity has topology $T^p \times S^1$ have the unusual
property that the entropy grows less than linearly in the mass.  It
might appear that it would therefore be entropically favorable for
these solutions to fragment into a number of smaller black holes of
the same type. This would constitute a new instability for these
solutions. This instability is also cause for concern, as we might be
able to violate the Bekenstein bound if we had enough small black
holes in a finite region.

For the BTZ black hole, there is an elegant proof that such an
instability is in fact impossible: any pair of black holes in an
asymptotically AdS$_3$ spacetime is always contained within a larger
black hole \cite{steif:multi-BTZ,brill:multi-BTZ}. In any attempt to
construct initial data describing a pair of black holes, if the
separation between them is small, the spatial section is closed and
there is no asymptotic region, while if the separation is larger,
there is extra energy from separating the black holes and the radius
of the resulting black hole is greater than the separation. This
answer accords well with our intuition about the Bekenstein bound;
whenever we try to violate the bound by packing a lot into a small
volume, we find that the energy is so large that the whole system
already lies inside a larger black hole.  

In general, for any asymptotically AdS$_{p+2}$ solution, we do not
expect to be able to separate black holes which are large compared to
the cosmological scale. What this means is that if we initially have a
pair of large separated black holes, we expect they will merge after a
time of order $b$, which is short compared with the characteristic
evolution time~$\sim \rho_0$ associated with the black holes. We
therefore cannot treat the black holes as separate thermodynamic
systems. In particular, we cannot apply the formula (\ref{tent}) for
the entropy of a static black hole in this case. Note that this also
explains why we should not be worried about such an instability for
large black holes in the case with spherical boundary conditions, even
though they also have an entropy which grows less than linearly in the
mass.

In the case of the higher-dimensional toroidal black holes, we still
have to worry about black holes with horizons smaller than the
cosmological scale. As $\rho_0 < b$, the characteristic evolution
timescale of the black holes is short, and the entropy should be
well-approximated by adding the entropies (\ref{tent}) for the
individual black holes. 

We should try to construct initial data corresponding to such
multi-black hole solutions. To simplify the problem, we assume that
the solution remains independent of all but one of the $y_i$; that is,
we just separate the black holes in say the $U,y_1$ plane. The general
form of the initial data is then
\begin{equation} \label{idat}
ds^2 = f(U,y_1) dU^2 + g(U,y_1) U^2 dy_1^2 + h(U,y_1) dy_i^2.
\end{equation}
Since we assume the solutions are independent of $p-1$ of the $y_i$,
we can eliminate these dimensions by Kaluza-Klein reduction to obtain
an equivalent three-dimensional problem. The $p+2$ dimensional black
holes have non-constant curvature, so this three-dimensional problem
is not equivalent to the BTZ case. From the three-dimensional point of
view, this is because these black hole solutions involve a non-trivial
value for the scalar field arising from the $dy_i^2$ part of the
metric.

If we consider the initial data for the black hole (\ref{tor}) coming
from the near-horizon limit of the D3-brane, then the $U, y_1$ part of
the metric is the initial data for the three-dimensional metric. This
two-dimensional surface has curvature
\begin{equation}
R = -{2 \over b^2} \left( 1 + {U_0^4 \over U^4} \right).
\end{equation}
In the special case $U_0=0$, the curvature is constant; the
three-dimensional solution obtained by reduction of (\ref{tor}) is
then the $M=0$ BTZ black hole. We might have expected to get
AdS$_3$ instead, as $U_0=0$ is supposed to be pure anti-de Sitter
space.  However, because we are taking toroidal boundary conditions at
infinity, the $U_0=0$ solution is actually AdS space with a discrete
set of identifications. In general, the non-constant part of the
curvature is important only near the horizon of the black hole. If we
consider a black hole which is much smaller than the cosmological
scale, the proper size of the $y_1$ direction becomes small compared
to the cosmological scale, signaling the presence of a black hole,
long before we get to the region where the curvature due to the black
hole becomes important. Thus, outside of a small region near the
horizon, the initial data surface looks like the initial data for the
$M=0$ BTZ black hole.

For initial data describing more than one small black hole, there is a
region away from the horizon, but still on scales small compared to
the cosmological scale, where we can argue that the metric looks like
the $M=0$ BTZ black hole. Thus, the full initial data surface (apart
from the small region around each horizon) should be well-approximated
by the multi-BTZ case. But we know from
\cite{steif:multi-BTZ,brill:multi-BTZ} that in that case, there is a
larger black hole horizon encompassing the others. Thus, it seems
reasonable to conjecture that for black holes small compared to the
cosmological scale in the higher-dimensional case, the same is true.
That is, we conjecture that there is no such instability for these
black holes either.

There is no contradiction between this conjecture and the fact that we
can separate branes. The near-horizon geometry of two groups of
conformal branes with a small separation between them is not a direct
product of the form AdS$_m \times$S$^n$, so it is not included in this
discussion, where we have been considering just the AdS part.  

\sect{Discussion}

Our main point is that the S$^{D-p-2}$ localization phase transition
observed in \cite{bdhm:adscft}, which was seen when the horizon size
gets down to the cosmological scale, does not occur for the spacetimes
which arise in the near-horizon limit of D$p$- or M-branes. This
transition arises for $S^p$ spatial boundary conditions, but the
near-horizon limit gives $T^p$ boundary conditions (or the
large-radius limit ${\bf R}^p$). Thus, this localization transition
does {\em not} imply an instability of the D$p$- or M-brane solutions,
in agreement with the general expectation that there is no such
instability.

We also considered a potential instability for AdS black holes to
break up into smaller black holes, and argued that it does not occur
either. In any initial data which describes several small black holes
in an asymptotically AdS spacetime, there should be a larger black
hole horizon which encompasses them. This is consistent with the fact
that a group of D$p$-branes can break up into smaller ones, because
the near-horizon geometry does not retain the simple direct product
form when the D$p$-branes break up.

For the case with $S^p$ spatial boundary, the black hole
correspondence principle tells us that there are further distinct
phases as the mass of the system gets even lower.  As discussed in
\cite{bdhm:adscft}, when the horizon size of the $D$ dimensional
Schwarzschild black hole gets down to the string scale, the system
goes into a Hagedorn, or long string, phase.  At still lower energies,
we see a gas of supergravitons. 

For the $T^p \times S^1$ case, there are further phases at energies
below that of the toroidal black hole, but the structure is rather
different, and neither the long string or AdS supergraviton gas phases
will appear.  For the $T^3$ case, the additional phases were analyzed
in \cite{Barbon:torDp}. The key is that there is a torus with a
$U$-dependent size. At low temperature, the horizon scale $\rho_0$ is
smaller than the string scale, so we must T-dualize at some $U>U_0$,
resulting in a ``smeared'' D0-brane spacetime.  Note that this
transition is at a gauge theory temperature which is higher than that
at which D3-brane finite-size effects kick in, taking into account
D3-brane fractionation \cite{Banks:matrix8dbh}.  At low enough
temperature, there is a localization to a D0-brane spacetime at
smaller $U$.  Note that in this phase, the $D=10$ black hole carries
Ramond-Ramond charge, unlike the $D=10$ Schwarzschild black hole which
arises for the $S^3$ boundary conditions.  At even lower temperatures,
there are further phases with more eleven-dimensional structure. For
the other $T^p$ cases, there should similarly be additional phases
which appear when $\rho_0$ is less than the relevant length scale.

{}From the gauge theory point of view, the localized black hole phase
is particularly interesting. This phase appears only for gauge theory
on the $p$-sphere at strong coupling. It breaks the $SO(6)$
R-symmetry, but without breaking spherical symmetry on the $S^p$. This
is similar to the Coulomb phase which appears for toroidal boundary
conditions, but the two are otherwise very different. The most
interesting aspect of the localized black hole phase is that the
Schwarzschild black hole has a negative specific heat. This is the
only instance of which we are aware in which a black hole with
negative specific heat is represented in the gauge theory.  It would
be interesting to investigate this further.

\vskip1truein
\centerline{\bf Acknowledgments}
\medskip

It is a pleasure to thank Gary Horowitz for discussions. {\small
A.W.P.}  wishes to thank the Aspen Center for Physics for hospitality
during early stages of this work.  The work of {\small S.F.R.} was
supported in part by NSF grant PHY95-07065, and that of {\small
A.W.P.}  by NSF grant PHY94-07194.

\newpage

\begingroup\raggedright\endgroup

\end{document}